\documentclass[letterpaper,twocolumn,10pt]{style/sig-alternate-10pt}

\usepackage{graphicx}

\usepackage[plainpages=false,pdfpagelabels=true,pdftitle={CAIR: Using Formal Languages to Study Routing, Leaking, and Interception in BGP},pdfauthor={Johann Schlamp, Matthias Waehlisch, Thomas C. Schmidt, Georg Carle, Ernst W. Biersack},bookmarks=true,bookmarksopen=true,pdfpagemode={UseOutlines},pdfstartview={FitV},pdfpagelayout={SinglePage},pdfborder={0 0 0},filecolor=black,linkcolor=black,urlcolor=black,draft]{hyperref}
\usepackage{breakurl}
\usepackage{booktabs}
\usepackage{multirow}
\usepackage{tabularx}
\usepackage{amsmath}
\usepackage{amssymb}
\usepackage{nccmath}
\usepackage[subrefformat=parens,labelformat=parens]{subfig}
\usepackage{dashbox}
\usepackage{style/tcolorbox}
\usepackage{microtype}
\usepackage{moresize}
\usepackage[normalem]{ulem}
\usepackage{accents}
\usepackage{enumitem}
\usepackage{cite}
\usepackage{balance}
\usepackage{xspace}

\definecolor{VictimGreen}{HTML}{789440}
\definecolor{AttackerRed}{HTML}{C05046}
\definecolor{NeutralBlue}{HTML}{0174DF}
\definecolor{egray}{rgb}{0.5,0.5,0.5}

\DeclareMathAlphabet{\mathbit}{OML}{cmr}{bx}{it}

\makeatletter
\def\paragraph{%
    \vskip 1pt\@startsection{paragraph}{4}{\z@}{2\p@ \@plus \p@}
    {-5\p@}{\textbf}%
}
\makeatother

\newcommand{\one}{({\em i})\xspace}
\newcommand{\two}{({\em ii})\xspace}
\newcommand{\three}{({\em iii})\xspace}

\let\orgautoref\autoref
\renewcommand{\autoref}
{\def\sectionautorefname{Section}%
\def\subsectionautorefname{Section}%
\def\subsubsectionautorefname{Subsection}%
\orgautoref}

\newcommand{\subfigref}[1]{\hyperref[#1]{Figure~\ref{#1}}}
\newcommand{\subtableref}[1]{\hyperref[#1]{Table \ref{#1}}}

\begin{document}
\date{}

\title{CAIR: \ Using Formal Languages to Study\\ Routing, Leaking, and Interception in BGP}


\numberofauthors{5}
\author{
\alignauthor Johann Schlamp\\
       \affaddr{Technische Universit\"at M\"unchen}\\
       \email{schlamp@in.tum.de}
\and
\alignauthor Matthias W\"ahlisch\\
       \affaddr{Freie Universit\"at Berlin}\\
       \email{m.waehlisch@fu-berlin.de}
\and
\alignauthor Thomas C. Schmidt\\
       \affaddr{HAW Hamburg}\\
       \email{t.schmidt@haw-hamburg.de}
\and
\alignauthor Georg Carle\\
       \affaddr{Technische Universit\"at M\"unchen}\\
       \email{carle@in.tum.de}
\and
\alignauthor Ernst W. Biersack\\
       \affaddr{Eurecom\\ Sophia Antipolis}\\
       \email{erbi@eurecom.fr}
}

\maketitle


\subsection*{Abstract}

The Internet routing protocol BGP expresses topological reachability and
policy-based decisions simultaneously in path vectors. A complete view on the
Internet backbone routing is given by the collection of all valid routes, which is
infeasible to obtain due to information hiding of BGP, the lack of omnipresent
collection points, and data complexity. Commonly, graph-based data models are
used to represent the Internet topology from a given set of BGP routing tables
but fall short of explaining policy contexts. As a consequence, routing
anomalies such as route leaks and interception attacks cannot be explained with
graphs.

In this paper, we use formal languages to represent the global routing system
in a rigorous model. Our CAIR framework translates BGP announcements into a
\textit{finite route language} that allows for the incremental construction of
minimal \textit{route automata}. CAIR preserves route diversity, is highly
efficient, and well-suited to monitor BGP path changes in real-time. We
formally derive implementable search patterns for route leaks and interception
attacks. In contrast to the state-of-the-art, we can detect these incidents. In
practical experiments, we analyze public BGP data over the last seven years.

\section{Introduction}

Measuring, modeling, and analyzing the routing between Autonomous Systems
(ASes) have gained increasing importance over the last decade
\cite{wr-itrr-13,motamedi:survey,df-itds-07}, as the Internet
infrastructure has become business critical and investigations of cyber
security incidents intensified. Today the question of route integrity is
prevalent in day-to-day operations and a clear understanding of how data should
flow is urgently desired to quickly detect anomalies. However, modeling
Internet routing is still a major challenge due to the complex decision making
in the Internet backbone. This work contributes a rigorous approach to route modeling
and analysis. 

The Internet inter-domain routing system selects those paths from the
topologically possible paths that are economically feasible and comply to
individual policies. Path vectors in BGP represent the outcome of this
hybrid decision process without revealing its underlying rules explicitly.
As such, BGP effectively hides most of its operational semantics from
observers and successfully withstands the quest for a simple explanatory
model \cite{rwmpb-lymmi-11}. The collection of all locally valid paths is
essentially what we can learn at any given observation point. These paths
are represented by our \textit{Constructible Automata for Internet Routes
(CAIR)}. CAIR offers two key advantages over existing solutions:

\begin{enumerate}[itemsep=0ex,topsep=0ex]
\item[a)] By preserving policy-related information in its routing model, CAIR can
reliably detect interception attacks or route leaks as they violate policies. 
\item[b)] CAIR is an efficient yet complete representation of the
observable inter-domain routing, which opens up the field for new analyses---even in
real-time.
\end{enumerate} 

Inter-provider connections are traditionally modeled as a
graph~\cite{motamedi:survey}. BGP peerings that show up in AS paths are
considered valid links between nodes (ASes). Network graphs adequately
represent connectivity in terms of router links or AS peerings. For the
analysis of policy-influenced routing, though, such graphs tend to oversimplify
the real situation. Realistic routing paths are selected on a per-prefix basis and often
influenced by local routing policies. In particular, network graphs falsely
imply transitivity over individual links and thus introduce additional,
potentially nonexistent (sub)paths. Therefore, graph models cannot
diagnose policy violations such as route leaking or 
complex
anomalies such as interception attacks.

In this paper, we solve the fundamental transitivity problem in graphs (\textbf{\S~\ref{sec:background}}) for Internet routes.
We remain with the concept of (context-dependent) path vectors
and introduce a \textit{finite route language} (\textbf{\S~\ref{sec:model}}) to model BGP data.
Starting from an alphabet of AS numbers and prefixes, we formally construct a
complete description of the observable routing system. With
corresponding \textit{route automata}~(\textbf{\S~\ref{sec:cair:automata}}), we show how to put this concept into
practical use. Using incrementally minimized finite-state automata (\textbf{\S~\ref{sec:practical}}), we arrive
at a most efficient representation of the path vector
space, which outperforms network graphs. Route automata are policy-aware: They represent the full characteristics
of the observable routing policies and, at the same time, reduce complexity to
real-time compliant processing. With CAIR, we can further derive
formal detection patterns for route anomalies and apply these to real BGP data.

In our analysis of real-world incidents (\textbf{\S~\ref{sec:evaluation}}), we focus
on two intricate anomalies. First, we search for interception attacks, in which an
adversary falsely attracts traffic while maintaining a backhaul path to its
victim to relay eavesdropped packets.  Second, we study (ab)normal routing
changes and gain insight into customer ASes that erroneously advertise transit
in a so-called \textit{route leak}. We validate our approach with well-known
incidents and identify 22 so far unknown cases of interception.

\section{Background and Intuition}\label{sec:background}

We approach CAIR by describing a gap we currently face
in network modeling and analysis in connection with common graph models. We then illustrate with some background why
a model based on formal languages and automata can actually close this gap. 
For detailed background on formal languages we refer to~\cite{hopcroft:intro}.

\subsection{Motivating Example: Interception Attacks}\label{sec:background:interception}

\begin{figure}[t!]
 \centering
 \subfloat[Regular traffic flow]{\label{fig:interception-forwarding-before}\includegraphics[width=0.22\textwidth]{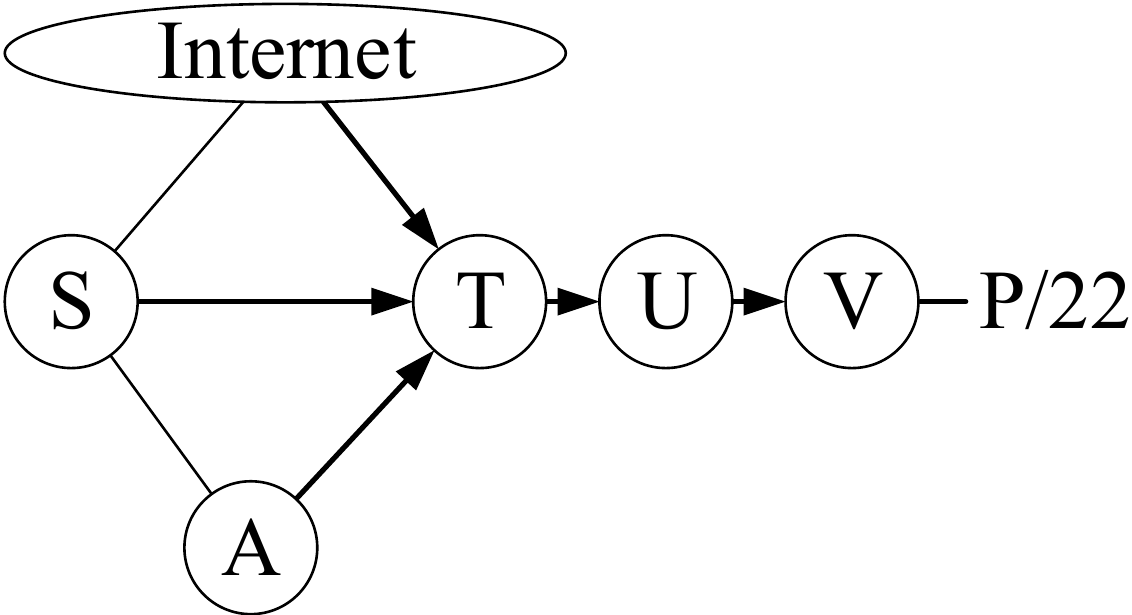}}
 \hfill
 \subfloat[Interception]{\label{fig:interception-forwarding-after}\includegraphics[width=0.24\textwidth]{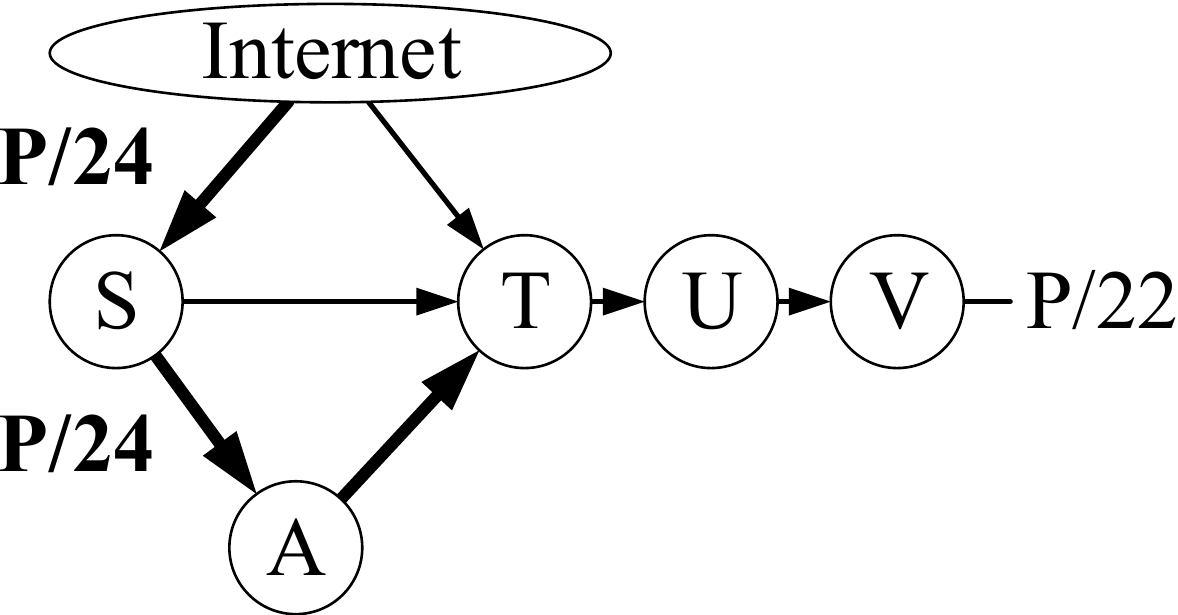}}
 \caption{Attacker $A$ intercepts subprefix $P/24$.}
 \label{fig:interception-forwarding}
\end{figure}

Interception attacks are characterized by a subtle injection of illegitimate
prefix routes to redirect traffic destined for a victim to the attacker.
To search for such events we need patterns and data models.
A crucial factor is to sustain global connectivity by relaying all communication
back to the victim. 
Implementing such an attack in the Internet is feasible and has been demonstrated in practice~\cite{defcon,hcg-clrac-13,renesys-china-hijack}.
A malicious AS needs to be connected to the Internet via at least two upstream ISPs.
The first ISP is used to attract traffic by advertising the victim's address space or a part thereof, also called a \textit{hijacking attack}.
The second ISP serves to preserve a stable \textit{backhaul path} from the attacker to the victim. 
To implement the backhaul path, the attacker includes all ASes between himself and his victim in the malicious BGP announcement.
As a consequence, all of these ASes discard the announcement due to loop prevention, and a stable link from the attacker to the victim's AS remains intact.
\autoref{fig:interception-forwarding} illustrates
the corresponding change in traffic forwarding.

Our key observation to derive a search pattern for interception attacks is that an arbitrary ISP $T$ normally redistributes \emph{all} routes of a \emph{distant} AS $V$ along the same ways \cite{gsg-sirp-13,anccg-iirpw-15} (\subfigref{fig:interception-forwarding-before}).
From a BGP policy point of view, however, the attack induces a specific policy for the redirected prefix. 
In our example, $S$ forwards traffic to $V$ via the malicious AS $A$ only for the prefix under attack, but reaches all other prefixes of the victim directly via $T$ (\subfigref{fig:interception-forwarding-after}). 
It is worth mentioning that the attacker can easily hide his own identity by not adding his AS number to the \texttt{AS\_PATH} attribute of forged BGP messages, as it is common practice with route servers~\cite{draft-ietf-idr-ix-bgp-route-server}, for instance.

To diagnose the difference, data models are needed to express route diversity.
Unfortunately, common graph models lack this capability.

\subsection{The Need for\hspace{5pt}{\texttt{\fontsize{14}{16}\selectfont CAIR}}}\label{sec:need-cair}

\begin{figure*}[t!]
 \centering
 \subfloat[Observed routes]{\label{fig:rib-example-list}\raisebox{.165\height}{\includegraphics[width=0.174\textwidth]{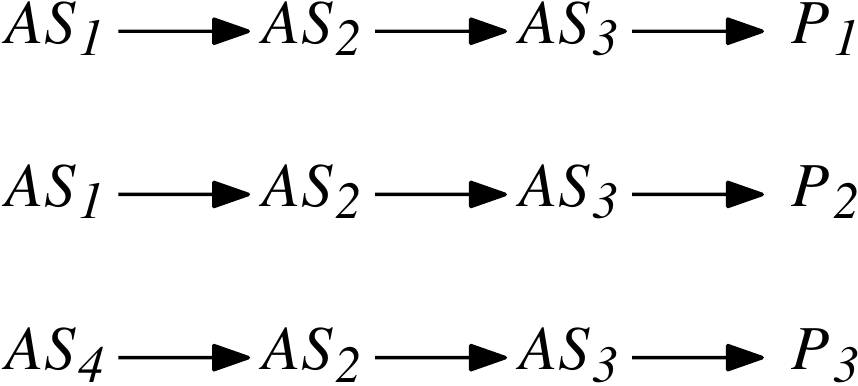}}}
 \hfill
 \subfloat[Network graph]{\label{fig:rib-example-graph}\raisebox{.097\height}{\includegraphics[width=0.155\textwidth]{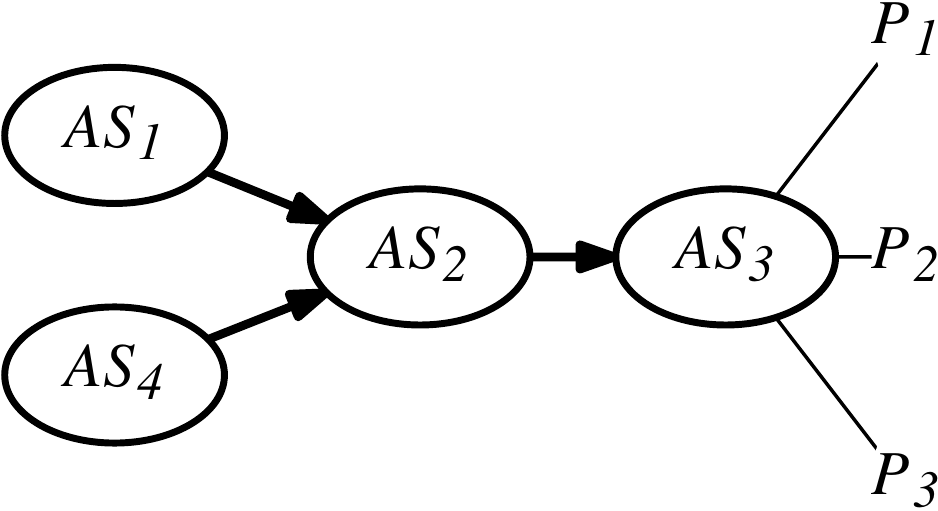}}}
 \hfill
 \subfloat[Deterministic automaton]{\label{fig:rib-example-dfa}\raisebox{.01\height}{\includegraphics[width=0.310\textwidth]{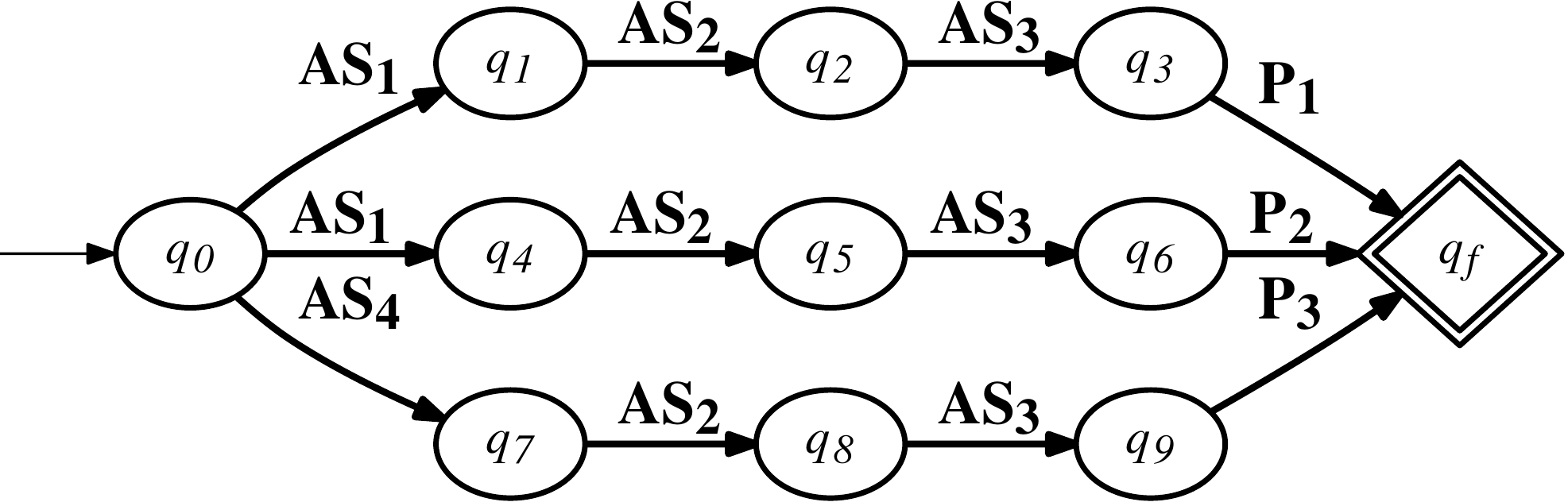}}}
 \hfill
 \subfloat[Minimal automaton]{\label{fig:rib-example-mdfa}\raisebox{.25\height}{\includegraphics[width=0.32\textwidth]{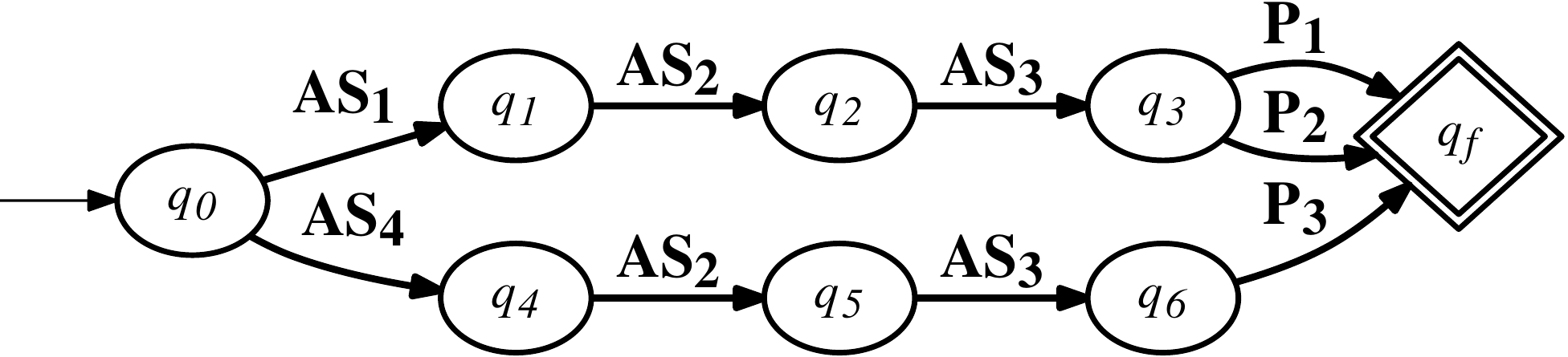}}}
 \caption{Representations of (re)distributed routes for prefixes $P_{1-3}$ by autonomous systems $AS_{1-4}$.}
 \label{fig:rib-example}
\end{figure*}

Consider a routing table extract that includes routes to three IP prefixes $P_{1-3}$
originated by the same AS $AS_{3}$ as shown in \subfigref{fig:rib-example-list}.
Obviously, a complete list of all paths preserves the observable routing
properties but is inefficient to memorize and difficult to analyze. A common
way to represent such network topologies reduces data to a graph $\mathcal{G} =
(N, L)$ that is a set of nodes $N$ and links $L$ describing connectivity
between ASes. For the analysis of policy-based routing, though, such
graphs tend to oversimplify real BGP operations. In our example, transit paths
to $AS_3$ differ for individual prefixes.
Such prefix-based policies~\cite{anccg-iirpw-15} are not reflected
in a graph model (compare to \subfigref{fig:rib-example-graph}). Rather, the model implies transitivity, which presumes
reachability that has not been observed. It is a modeling artifact.

\medskip

\noindent\fbox{\parbox[c][3.75em][c]{0.97\columnwidth}{
\medskip
\hspace{1.25ex}\textbf{Incorrect Transitivity Assumption in Graphs:}
$$
AS_1 \rightarrow AS_2 \ , \ AS_2 \rightarrow AS_3 \ \Rightarrow \ AS_1 \rightarrow AS_3
$$
}
}

\medskip

Note that in our example, transitivity breaks for the prefix $P_3$, i.e.~no route
$AS_1 \rightarrow AS_2 \rightarrow AS_3 \rightarrow P_3$ has been observed.
CAIR offers a natural solution to this problem. It allows for an accurate
representation of observed routing paths and is a highly efficient
approach at the same time.

\subsection{Why Using Finite Automata}

A finite language consists of characters and rules to construct words, which
can be represented by a deterministic finite-state automaton (DFA). In this
paper, we understand a given set of network paths as such a finite language. AS
paths towards prefixes represent words of this language. Our CAIR framework
serves to construct and minimize corresponding automata providing the following
unique features:

\begin{description}[leftmargin=9pt,itemsep=0ex]
  \item[Accuracy] All initially observed BGP paths are stored in an automaton
without introducing further unobserved information. Consequently, CAIR
accurately reflects the observable routing system.
  \item[Expressiveness] The states of a minimal DFA represent equivalence
classes such that two equivalent states exhibit the same (routing) behaviour.
CAIR exploits this unique fact to reveal intrinsic routing properties, such as
the importance of an AS or the location of a hidden attacker.
  \item[Feasability] Well-understood algorithms exist to construct a DFA.
Processing of paths equals traversing the automaton. As a consequence, CAIR
supports random access of data. It can be deployed both in real-time to
continuously monitor BGP as well as in retrospective on archived data sets.
  \item[Efficiency] Any DFA can be minimized without sacrificing accuracy. The
resulting minimal automaton is unique. This is a highly efficient way to
represent a finite language in general \cite{hopcroft:intro}, and consequently a set of network paths
we model with CAIR.
\end{description}

Despite this powerful feature set, concepts of formal languages have not been
applied, yet, in the context of Internet routes. It is worth mentioning that
other, less popular data structures could correctly model network paths,
e.g.~\textit{tries}~\cite{briandais:trie}. We will see later, however, that
CAIR naturally outperforms such data structures in terms of expressiveness and
efficiency.

\subsection{{\texttt{\fontsize{14}{16}\selectfont CAIR}}\hspace{5pt}in a Nutshell}

CAIR derives a \textit{finite route language} and constructs a \textit{route automaton} based on BGP data such that each transition (edge) is labeled with either an AS number or an IP prefix. CAIR thereby reduces the
amount of states by on-the-fly minimization. Traversing edges until the final,
accepting state of the automaton results into a complete AS path as included in
the initial BGP data set. Having the automaton in place, we search for specific
properties. Since a DFA exists for any finite language (and vice versa), route
automata accurately model the underlying input data and can thus be used to
precisely describe its intrinsic routing characteristics.

\paragraph*{Toy Model}

For our example in \subfigref{fig:rib-example-list}, corresponding DFA and
minimal DFA (MDFA) are shown in \subfigref{fig:rib-example-dfa} and
\subfigref{fig:rib-example-mdfa} respectively. In our CAIR framework, we
utilize the latter one for route analysis. An important aspect of
minimization is that we can observe route diversity based on nonminimizable states.
In the minimal automaton, we obtain a sequence of independent states $(q_1, q_2, q_3)$
and $(q_4, q_5, q_6)$ that represents the nonuniform
redistribution of prefixes $\{P_1, P_2\}$ and $\{P_3\}$ (compare
\subfigref{fig:rib-example-dfa} and \subfigref{fig:rib-example-mdfa}).

It is worth noting that CAIR does not intend to improve on the incompleteness of
measured BGP data. Instead, we present a novel data model with maximum expressiveness
that allows for an accurate representation of \emph{observed} routing paths.

\section{Finite Route Languages}\label{sec:model}

A formal language is a set of strings of symbols constrained
by specific rules. The global Internet routing system can be 
 represented as all active BGP routes,~i.e.~a set of
AS paths from all vantage points towards all advertised IP prefixes.
By adopting the immediate analogy, we define the \textit{Finite Route 
Language (FRL)} as follows.

\paragraph*{Definition}

Let $\Sigma_{AS}$ be the set of all ASes, $\Pi$ the set of all IP addresses, $p
\subset \Pi$ an IP prefix, and $p' \subset p$ a more specific prefix of $p$.
Let further be $(w,p) \in \Sigma_{AS}^* \times \Pi$ a route with an AS path $w
\in \Sigma_{AS}^*$, i.e.~an arbitrary concatenation of ASes, to a prefix $p \in
\Pi$, in the following denoted $r = wp$. Then, we define $\mathcal{L} \subset
\Sigma_{AS}^* \times \Pi$ as the set of \textit{active} routes to all
advertised prefixes in the global routing system, i.e.~the set of all
observable routes. $\mathcal{L}(p) \subset \mathcal{L}$ denotes the subset of
routes to a given prefix $p \subset \Pi$, such that $$\mathcal{L}(p) \ = \
\{wuop \in \mathcal{L} \ | \ w \in \Sigma_{AS}^* \ ; \ u \in \Sigma_{AS} \ ; \
o \in \Sigma_{AS}\}$$ with $w$ being an AS subpath and $u$ the upstream AS of
the origin AS $o$. For a given route $r$ and a subprefix $p' \subset p$, we
postulate $r \in \mathcal{L}(p) \Rightarrow r \in \mathcal{L}(p')$ as a
corollary, since routes to less specific prefixes also cover more specific
prefixes. Note that the converse is false. Further, $\mathcal{L}_{P} \subset
\mathcal{L}$ denotes the set of all observable routes from a set of observation
points $P \subset \Sigma_{AS}$. We reuse the unary operator $|\ .\ |$ to
indicate the number of routes in a set $\mathcal{O} \subseteq \mathcal{L}$, the
length of a route $r \in \mathcal{L}$ or a subpath $w \in \Sigma_{AS}^*$, and
the number of ASes in a set $S \subseteq \Sigma_{AS}$.

\subsection{Formalization of Routing Attacks}\label{sec:model:attacks}

BGP-based routing attacks are caused by injecting illegitimate
route updates that alter the global BGP routing table in order to 
modify traffic flows. Route injections are considered illegitimate, if they 
violate topological constraints or policies as derived from business relations.

\paragraph*{Attacker Model}

Our attacker is assumed capable of injecting arbitrary BGP messages into the
global routing system, i.e.~he operates a BGP router and maintains a BGP
session to at least one upstream provider. The attacker is not hindered by
local filters or other validation mechanisms of his upstream. Instead, the
upstream AS indifferently redistributes all update messages to its peers, which
thus may propagate throughout the Internet. An observation point shall be in
place to monitor the propagation of BGP messages. It is worth mentioning that
route updates with less attractive paths may not reach a particular observation
point due to \textit{best path selection} in BGP. Without loss of generality, an
omnipresent observation point to observe the set of all active routes
$\mathcal{L}$ is assumed for the following definitions.

\paragraph*{Generic Routing Attacks}

We denote an attacker's AS $a \in \Sigma_{AS}$ and his
victim's AS $v \in \Sigma_{AS}$. Further, a victim's prefix is given by
$p_{v} \subset \Pi$. Then, a generic routing attack is defined by an
attacker extending the set of valid routes $\mathcal{L}$ by injecting forged
routes $\mathcal{F}_{a}$ into the routing system, such that the altered set of
globally active routes $\hat{\mathcal{L}}(p'_{v})$ is given by
$$\hat{\mathcal{L}}(p'_{v}) \ = \ \mathcal{L}(p_{v}) \ \cup \ \mathcal{F}_{a}(p'_v) \text{\ \ with \ } p'_v \subseteq p_{v} \ .$$
In general terms, the attacker advertises reachability of a victim's
network---either entirely or in parts---and may thus attract a fraction of the
victim's inbound traffic.

\paragraph*{Requirements and Impact}

In BGP, the impact of a hijacking attack generally depends on a best
path selection process. In particular, shorter AS paths are preferred over
longer ones, although policy-induced exceptions on a case-by-case basis may
exist. With respect to packet forwarding, routes to longer IP prefixes prevail.
Assuming the ambition to forge globally accepted routes, an attacker thus
succeeds if his routes towards a victim's network are considered best by a vast
majority of Internet participants. This implies that an attacker needs to
ensure that his bogus routes $\mathcal{F}_{a}(p'_v)$ are either
\begin{itemize}
\item[1)] shortest from a global perspective, i.e.\newline$\forall r \in \mathcal{L}(p'_v), \ r_{a} \in \mathcal{F}_{a}(p'_v) \ \colon \ |r_{a}| < |r|$, or
\item[2)] more specific than all others, i.e.\newline$\forall p''_v \subseteq p'_v \subset p_v \ \colon \ \mathcal{L}(p''_v) = \mathcal{L}(p'_v) = \mathcal{L}(p_v)$.
\end{itemize}
As a consequence, the prospects of identifying an attack naturally depend on
the significance of changes in $\hat{\mathcal{L}}$ and on the peculiarities of
the forged routes.

\subsection{Application to Interception Attacks}\label{sec:model:interception}

A BGP-based interception attack succeeds if data packets sent to a victim
 are re-routed via an attacker's AS, who in turn needs to have a stable
backhaul path to the victim to forward the eavesdropped packets. Such attacks
are unfeasible to detect from a topological point of view since only new paths
over already existing links emerge. In the following, we use our formalized
route model to thoroughly study these subtle changes.

\paragraph*{Formal Model}

The attacker $a \in \Sigma_{AS}$ chooses to advertise (a part of) the victim's prefix $p_{v} \in \Pi$, while including a backhaul path in his announcemennts.
As a BGP speaker, the attacker can easily identify such a path to $p_v$ in
$\mathcal{L}_{a}(p_{v})$.
The set of AS paths that represent a suitable backhaul link via
an arbitrary upstream AS $u \in \Sigma_{AS}$ thus reads
$$ W^{a}_{v}(u) \ = \ \{ w^{a}_{v} \in \Sigma_{AS}^{*} \ | \ uw^{a}_{v}p_{v} \in \mathcal{L}_{a}(p_{v}) \} \ .$$
Note that in practice, exactly one such path $w^{a}_{v}$ exists per upstream AS,
since each AS redistributes its prefered routes only. For the attack
to succeed, the attacker needs to employ at least two upstream providers. In
the following, $t \in \Sigma_{AS}$ serves to establish the backhaul
link, while the remaining upstream providers $S \subset \Sigma_{AS}$ are
utilized to launch a particular (sub)prefix hijacking attack~\cite{heap} against $p'_{v}
\subseteq p_{v}$ such that
\begin{align}
\hat{\mathcal{L}}(p'_{v}) \ =& \ \ \{wvp_{v} \ | \ w \in \Sigma_{AS}^* \ \} \ \cup \ & & \text{\it\small legitimate} \notag \\
& \ \ \{wsatw^{a}_{v}p'_{v} \ | \ w \in \Sigma_{AS}^* \ ; \ & & \medmath{\mathbf{forged}} \notag \\
& \ \ \ \ \quad s \in S \ ; \ w^{a}_{v} \in W^{a}_{v}(t)\} \ . \notag
\end{align}
Most notably, the attacker can hide his AS number (see \autoref{sec:background:interception}) while still forwarding between the upstreams $s$ and $t$. In any case, the attack does not lead to suspicious topological changes. The observable sets of both origin and upstream ASes for $v$ remain unchanged,
while the forged route to $p'_{v}$ generates the impression of originating 
from the victim's AS $v$ and propagating via his legitimate upstream providers (as well as theirs).

\section{The\hspace{5pt}{\texttt{\fontsize{14}{16}\selectfont CAIR}}\hspace{5pt}Framework}\label{sec:cair:automata}

After introducing the concept of
\emph{Constructible Automata for Internet Routes (CAIR)}, we 
formulate an implementable search pattern for interception attacks.

\subsection{Route Automata}\label{sec:cair:automata:notation}

We define a \textit{route automaton} as a \textit{minimal} deterministic finite-state
automaton that accepts any given finite route language (see
\autoref{sec:model}).
In general terms, an automaton represents a state machine that accepts a formal
language, i.e.~processing one of its words ends in an accepting state. Let
$\mathcal{L} \subset \Sigma_{AS}^* \times \Pi$ be an FRL
representing all routes in
the global routing system. Then, we define a route automaton as the 5-tuple
$$M \ = \ (Q, \ \Sigma_{AS} \cup \Pi, \ \delta, \ q_0, \ F)$$
with $Q$ a finite set of states, $q_0 \in Q$ the start state, $F \subset Q$ a
set of accepting states, and $\delta$ a partial mapping $\delta: Q \times
\Sigma_{AS} \cup \Pi \rightarrow Q$ denoting transitions. We define the
extended transition function $\delta^*$ for routes
$u\vec{r} \in \mathcal{L}$ as the
mapping $\delta^*: Q \times \mathcal{L} \rightarrow Q$ such that
\begin{align*}
\delta^*(q,\epsilon) & \ = \ q \\
\delta^*(q,u\vec{r}) & \ = \ \begin{cases} \delta^*(\delta(q,u),\vec{r}) & \text{if \ } \delta(q,u) \ne \bot, \\ \bot & \text{otherwise} \end{cases}
\end{align*}
with $u \in \Sigma, \ \vec{r} \in \Sigma^* \times \Pi$ a partial route, $q \in Q, \ \epsilon$ an
empty path, and $\bot$ a catching state with $\bot \notin Q$ that represents nonexistent routes. We further
define $\mathcal{L}(M)$ as the language accepted by the automaton $M$ as
$$\mathcal{L}(M) \ = \ \{r \in \mathcal{L} \ | \ \delta^*(q_0, r) \in F \} \ .$$
Then, $M_P$ denotes an automaton accepting a set of observed routes
$\mathcal{L}_P \subset \mathcal{L}$ from individual observation points $P
\subset \Sigma_{AS}$ such that $\mathcal{L}(M_P) = \mathcal{L}_P$.

A path segment $w \in \Sigma_{AS}^*$ with $r =
w\vec{r} \in \mathcal{L}, \ |w| < |r|$ is denoted
$w \prec r$. We define the
longest common path segment $w_{lp}$ in $M$ such that
$$\forall r \in \mathcal{L}(M), \ w \prec r \ : \ |w_{lp}| > |w|, \ \delta^*(q_0,w_{lp}) \ \ne \ \bot \ .$$
The power set $\mathcal{P}(\mathcal{L}) $ holds all finite languages that
represent (partial) routes.  We define the \textit{right language} of a state
$q \in Q$ in $M$ as
$$\vec{\mathcal{L}}(q) \ = \ \{ \vec{r} \in \mathcal{P}(\mathcal{L}) \ | \ \delta^*(q,\vec{r}) \in F \} \ .$$
In other words, $\vec{\mathcal{L}}: Q \rightarrow \mathcal{P}(\mathcal{L})$
maps the state $q$ to the set of partial routes producible in the automaton $M$
starting at $q$. Note that the language accepted by $M$ can accordingly be
defined by $\mathcal{L}(M) = \vec{\mathcal{L}}(q_0)$.

All states $q \in Q$ that accept the same right language are equivalent and can
be merged into a single state. Iteratively applied, this process leads to a
minimal automaton, in the following called \textit{route automaton}. This
automaton is unique except isomorphisms and requires a minimum number of states
among all automata that accept the same language ~\cite{hopcroft:intro}. Note
that the number of accepting states in a route automaton is $|F| = 1$, since
all routes $r \in \mathcal{L}$ end with an IP prefix, i.e. $\forall q \in Q, \
p \in \Pi \ : \ \delta(q, p) \in F$ and $\forall q \in Q, \ w \in \Sigma_{AS}^*
\ : \ \delta^*(q, w) \notin F$. In the following, we use $q_f$ to refer to this
single accepting state.

\subsection{A Search Pattern for Interception Attacks}\label{sec:cair:detection}

\begin{figure}[t!]
  \centering
  \includegraphics[width=0.475\textwidth]{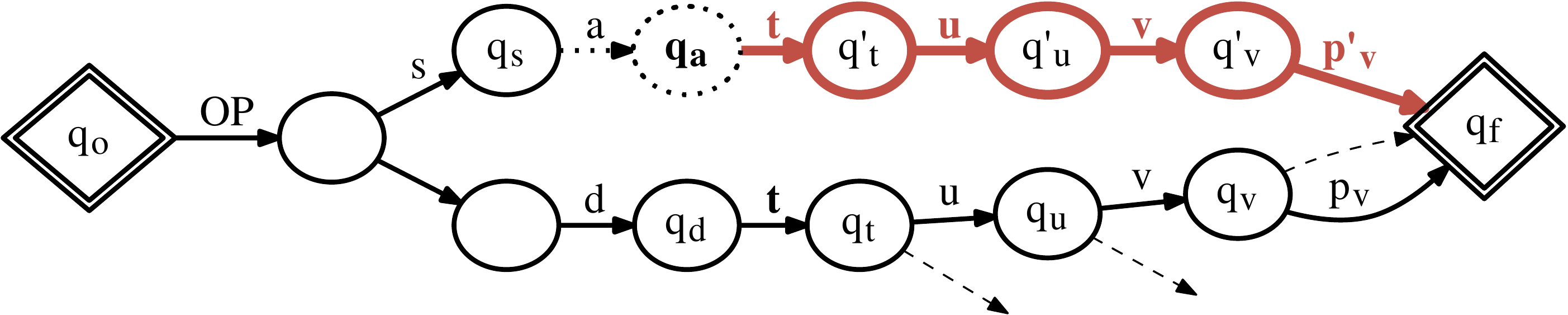}
 \caption{Route automaton for an interception attack.}
 \label{fig:cair_signature}
\end{figure}

We present a methodology to search route automata for
anomalies emerging from interception attacks. Before we explain the details,
we briefly illustrate the intuition behind our approach (for background
see \autoref{sec:background:interception}).

\subsubsection{Intuition behind the Detection Scheme}

\autoref{fig:cair_signature} shows an interception attack on a (sub)prefix
$p'_v \subseteq p_v \in \Pi$ of the victim $v \in \Sigma_{AS}$, in which the
attacker $a \in \Sigma_{AS}$ fabricates an \textit{artificial path segment} $w^a_v = tuv
\in \Sigma_{AS}^*$ over $t,u \in \Sigma_{AS}$ as described in
\autoref{sec:model:interception}. Recall our assumption that a distant AS~$t$
that is not a direct upstream of $v$ neutrally redistributes the routes of $v$
(\autoref{sec:background:interception}). Hence, the right language
(\autoref{sec:cair:automata:notation}) of the automaton state $q_t$
should represent all transit routes over $t$ (indicated by dashed lines in
\autoref{fig:cair_signature}). In an interception scenario, however, the
attacker seemingly changes the routing policy of~$t$: It appears that $t$ now
forwards announcements of $p_v$ and $p'_v$ differently,
i.e.~selectively to $a, d \in \Sigma_{AS}$. As a result of this diversity, our
route automaton holds separate states $\{q_v, q'_v\}$, $\{q_u, q'_u\}$, and $\{q_t, q'_t\}$.

In the remainder, we anticipate that the attacker hides his true identity
by not adding his AS number~$a$ to the \texttt{AS\_PATH} attribute of forged
BGP messages (remove dotted elements in \autoref{fig:cair_signature}). In this
case, we rename $q_s$ to $q_a$ with $s \in \Sigma_{AS}$ being the attacker's second
upstream provider. We further assume that the attacker seeks to
globally attract traffic to $v$ by means of a strict subprefix hijacking,
i.e.~$p'_v \subset p_v \in \Pi, \ p'_v \neq p_v$ holds.

\subsubsection{Translation to Route Automata}

In order to detect interception attacks, we need to search for a partial route
that a) is used in a single routing context only (\textit{artificiality}), b)
is in contradiction to existing announcements (\textit{nonuniformity}), and c)
leads to a subprefix of the benign routes (\textit{interception alert}).
Within our route automata, these conditions can be easily expressed and
implemented.

\paragraph*{Artificiality}

An artificial path segment $w^a_v \in \Sigma_{AS}^*$ is given by a sequence of (at least) four
states $\delta^*(q_a, tw^a_v) = q'_v$ with a single outgoing transition each, i.e.
$$\forall tw \in \Sigma_{AS}^*, \ w \neq w^a_v \ \colon \ \delta^*(q_a, tw) = \bot \ .$$

\paragraph*{Nonuniformity}

We further search for a path segment $w^d_v$ that is in contradiction with
$w^a_v$ such that
$$\exists q_d \in Q \ \colon \ \delta^*(q_d, w^d_v) = q_v \ .$$

\paragraph*{Interception alert}

We verify if the offending state $q'_v$ represents a subprefix hijacking of $q_v$, i.e.~if the condition
$$\exists p_v,p'_v \in \Pi, \ p'_v \subset p_v \ \colon \ \delta(q_v, p_v) = \delta(q'_v, p'_v) = q_f$$
holds true. We then raise an interception alert and report $p_v$, $p'_v$, and $w^a_v$ for
further investigation.

\subsubsection{Discrimination of the Attacker}\label{sec:cair:detection:loc}

With our route automaton, we are able to pinpoint the attacker's location in
the Internet topology. If the attacker adds his AS number $a$ to the BGP
updates, we can directly observe a transition labeled with $a$ that
points to the offending state $q_a$. Under the assumption that the attacker is
taking precautionary measures to hide his AS (refer to \autoref{sec:background:interception}), the incoming and outgoing
transitions of $q_a$ are labeled with $s$ and $t$ respectively. In this
case, we can isolate the attacker to be a customer of both ISPs $s$
and $t$, which leaves us with a small number of possible ASes to be 
manually~scrutinized.

\section{Incident Detection with\hspace{5pt}{\texttt{\fontsize{14}{16}\selectfont CAIR}}\hspace{5pt}}\label{sec:evaluation}

We now deploy CAIR in practice and analyze seven years of BGP data with respect
to two applications of CAIR, the detection of interception in BGP and route
leak analysis.  First, we apply our detection scheme and reveal 22~critical
interception incidents so far unknown with victims mostly belong to the R\&D sector and to the
medium-sized ISP business. We explain details along two case studies, including
our ground truth DEFCON~\cite{defcon}. Second, we derive measures to study the
importance of ASes with respect to global routing, and evaluate their
characteristics over time. We compare our results to a sudden and radical
routing change during a recent route leak of Telekom Malaysia
(2015)~\cite{malaysia}.

\subsection{BGP Interception Incidents}\label{sec:eval:application}

We use CAIR to construct route automata based on BGP routing table entries from
the RouteViews~\cite{routeviews} \textit{Oregon2} collector, August, 2008 --
November, 2015, analyzed in an interval of two weeks (i.e.~174~RIBs
with a total of $\approx$2.4B~entries).

\subsubsection{Overview}

In total, we identified 6,171 artificial path segments $w^a_v \in
\Sigma_{AS}^*$ (see \autoref{sec:model:interception}). For 527 of them, we
found a corresponding segment $w^d_v \in \Sigma_{AS}^*$ that evidenced
nonuniform redistribution of BGP updates
(\autoref{sec:background:interception}). Our detection scheme finally raised 41
alerts for (strict) subprefix hijacking attacks.
\autoref{table:cair_results:alerts} presents the results. Note that one of
these events lasted for 8 weeks, i.e.~accounts for four subsequent alerts.
Additional five events were observed twice. This leaves us with a total number
of 32 distinct cases.

\paragraph*{Sanitizing: Exclude false positives}

Our search pattern for interception is based on the observation that nonuniform
redistribution of BGP announcements should only take place at a victim $v$
itself or at his direct upstream ISPs, i.e.~we search for artificial path
segments of length $|w^a_v| \ge 3$. This assumption breaks if the victim
operates multiple ASes \cite{chkw-tam-10} that are consecutively visible in an
AS path. We manually analyzed the 32~interception alerts carefully, and were
able to identify 9 sibling cases, which we exclude from further investigation.
The remaining 23~alerts are marked as suspicious (\autoref{table:cair:attacks}).

\begin{table}[t!]
\small
\setlength{\tabcolsep}{4pt}
\renewcommand{\arraystretch}{1.14}
\begin{tabularx}{0.475\textwidth}{Xrrr}
  \ssmall August 10, 2008 -- November 1, 2015 & total & in \% \\
  \toprule
  \bf Artificial path segments \textcolor{egray}{$w^a_v$} & \bf 6,171 & \bf 100.0\% \\
  \midrule
  Nonuniform redistribution \textcolor{egray}{$w^{z}_v \ne w^a_v$} & 527 & 8.54\% \\
  Interception alerts$^1$ \textcolor{egray}{$p'_v \subset p_v$} & 41 & 0.66\% \\
  \bf Unique alerts (victims) & \bf 32 & \bf 0.52\% \\
  \midrule
  \it Manual inspection & & \\
  \midrule
  Sibling ASes (victim) & -7 & 0.11\% \\
  Sibling ASes (upstream) & -2 & 0.03\% \\
  \midrule
  \bf Alerts after manual inspection & \bf 23 & \bf 0.37\% \\
  \bottomrule
\end{tabularx}
\caption{Interception alerts raised by CAIR.}
\vspace{3pt}
\label{table:cair_results:alerts}
\centering
{\small$^1$Concerning subprefix announcements only.\vspace{6pt}}
\end{table}

\paragraph*{Substantial alerts}

We summarize the 23~alerts in \autoref{table:cair:attacks}.
Entries that are highlighted in grey represent the upstream AS $s \in \Sigma_{AS}$ that is used by an attacker
to redistribute forged BGP updates. The corresponding AS neighbor $t \in \Sigma_{AS}$
represents the attacker's second upstream, which is used to uphold the
backhaul path. The underlined parts of the AS paths show the
artificial segment $w^a_v$. In 13 cases, $w^a_v$ is of length 3 (plus the
attacker's two upstreams and possibly further intermediate ASes), while 6 cases exhibit a segment length of 4 AS hops. The 4 remaining
incidents show longer segments. To better grasp the details of the incidents, we checked
AS names in the WHOIS system. Note that some of these events date far back to
the past. We consequently utilized archived WHOIS data for the respective
dates.

\begin{table*}[t!]
 \setlength{\tabcolsep}{2.5pt}
 \renewcommand{\arraystretch}{0.85}
   \small
   \centering
   \begin{tabularx}{0.9\textwidth}{r|rrrrrrX}
     \toprule
     \multicolumn{1}{c|}{\bf Date} & \multicolumn{5}{l}{\textbf{\, Forged AS path} / nonuniform route segment} & \multicolumn{2}{l}{\textbf{\, Victim} / country / company} \\[1pt]
     \toprule
     $^{1,2}$2008/08/10 & & \textcolor{egray}{AS26627} & \uwave{AS4436} & \underline{AS22822} & \underline{AS23005} & \underline{AS20195} & United States, \textit{Sparkplug Las Vegas} \\
     \midrule
     2009/02/01 & \textcolor{egray}{AS3303} & \uwave{AS1299} & \underline{AS701} & \underline{AS3491} & \underline{AS37004} & \underline{AS30988} & Nigeria, \textit{IS InternetSolutions} \\
     2009/10/15 & \textcolor{egray}{AS3561} & \uwave{AS7018} & \underline{AS4837} & \underline{AS4808} & \underline{AS17431} & \underline{AS17964} & China, \textit{Beijing Network Technologies} \\
     \midrule
     2010/05/15 & \textcolor{egray}{AS2914} & \uwave{AS3549} & \underline{AS3356} & \underline{AS23148} & \underline{AS20080} & \underline{AS1916} & Brasil, \textit{Rede Nacional de Ensino e Pesquisa} \\
     2010/12/15 & & \textcolor{egray}{AS34984} & \uwave{AS12301} & \underline{AS3549} & \underline{AS9121} & \underline{AS12794} & Turkey, \textit{Akbank} \\
     \midrule
     2011/01/15 & \textcolor{egray}{AS9002} & \uwave{AS21230} & ..(+2).. & \underline{AS3356} & \underline{AS9121} & \underline{AS44565} & Turkey, \textit{Vital Teknoloji} \\
     2011/03/01 & \textcolor{egray}{AS4134} & \uwave{AS40633} & ..(+4).. & \underline{AS6453} & \underline{AS9299} & \underline{AS18223} & India, \textit{Capital IQ Information Systems} \\
     2011/04/01 & & \textcolor{egray}{AS3549} & \uwave{AS5391} & \underline{AS25144} & \underline{AS42432} & \underline{AS8670} & Bosnia, \textit{University of Sarajevo} \\
     2011/04/01 & & \textcolor{egray}{AS3549} & \uwave{AS10026} & \underline{AS9957} & \underline{AS10036} & \underline{AS18334} & South Korea, \textit{Gyounggidongbu Cable TV} \\
     2011/08/15 & & \textcolor{egray}{AS1273} & \uwave{AS1299} & \underline{AS3491} & \underline{AS20485} & \underline{AS8402} & Russia, \textit{Vimpelcom} \\
     2011/12/01 & & \textcolor{egray}{AS6762} & \uwave{AS31133} & \underline{AS12695} & \underline{AS34123} & \underline{AS28738} & Russia, \textit{InterLAN Communications} \\
     2011/12/15 & & \textcolor{egray}{AS702} & \uwave{AS701} & \underline{AS3549} & \underline{AS21371} & \underline{AS49669} & United Kingdom, \textit{Cognito} \\
     \midrule
     2012/09/15 & & \textcolor{egray}{AS2828} & \uwave{AS1299} & \underline{AS9498} & \underline{AS58459} & \underline{AS4613} & Nepal, \textit{Mercantile Office Systems} \\
     2012/11/01 & \textcolor{egray}{AS30496} & \uwave{AS11427} & AS7843 & \underline{AS6461} & \underline{AS33481} & \underline{AS40610} & United States, \textit{Digital Passage} \\
     2012/12/15 & \textcolor{egray}{AS30496} & \uwave{AS11427} & AS7843 & \underline{AS6461} & \underline{AS33481} & \underline{AS21854} & United States, \textit{Digital Passage} \\
     \midrule
     2013/02/15 & & \textcolor{egray}{AS1273} & \uwave{AS2914} & \underline{AS8928} & \underline{AS5391} & \underline{AS57888} & Croatia, \textit{Telesat} \\
     2013/06/01 & \textcolor{egray}{AS1299} & \uwave{AS6663} & ..(+3).. & \underline{AS6939} & \underline{AS197043} & \underline{AS197890} & Germany, \textit{Megaservers} \\
     2013/07/15 & & \textcolor{egray}{AS3356} & \uwave{AS1299} & \underline{AS6663} & \underline{AS41571} & \underline{AS48828} & Romania, \textit{Carosystem} \\
     $^2$2013/08/15 & \textcolor{egray}{AS3549} & \uwave{AS3491} & \underline{AS12880} & \underline{AS43343} & \underline{AS21341} & \underline{AS25306} & Iran, \textit{Institute IsIran} \\
     2013/12/01 & & \textcolor{egray}{AS1299} & \uwave{AS9498} & \underline{AS12880} & \underline{AS41881} & \underline{AS51411} & Iran, \textit{Toos-Ashena} \\
     \midrule
     2015/02/15 & & \textcolor{egray}{AS1299} & \uwave{AS52320} & \underline{AS16735} & \underline{AS28284} & \underline{AS262353} & Brasil, \textit{Marcelo Bonini} \\
     2015/07/15 & \textcolor{egray}{AS3356} & \uwave{AS209} & ..(+2).. & \underline{AS721} & \underline{AS27066} & \underline{AS747} & United States, \textit{US Army ISC} \\
     2015/08/15 & & \textcolor{egray}{AS46450} & \uwave{AS1299} & \underline{AS3356} & \underline{AS6079} & \underline{AS55079} & United States, \textit{Third Gear Networks} \\[1pt]
     \bottomrule
     \multicolumn{8}{c}{}\\[-4pt]
     \multicolumn{1}{c}{} & \multicolumn{6}{c}{\small $s$: ~\textcolor{egray}{ASN} \quad\quad $t$: ~\uwave{ASN} \quad\quad $w^a_v$: ~\underline{ASN}~...~\underline{ASN}} \\
   \end{tabularx}
   \caption{Remaining interception alerts after manual inspection.}
   \label{table:cair:attacks}
   \centering
   {\small$^1$Public demonstration of an interception attack at DEFCON~\cite{defcon}.\\
    $^2$Studied more closely in \autoref{sec:eval:application:cases}.}
\end{table*}

\subsubsection{Case Studies}\label{sec:eval:application:cases}

\paragraph*{Ground Truth: The DEFCON Attack (2008)}

The DEFCON attack~\cite{defcon} was a well-known experiment to demonstrate
interception attacks in the Internet. All details on the attack are publicly
available~\cite{blackhat} and thus provide perfect ground truth for validating
CAIR.

\begin{figure*}[t!]
 \centering
 \includegraphics[width=\textwidth]{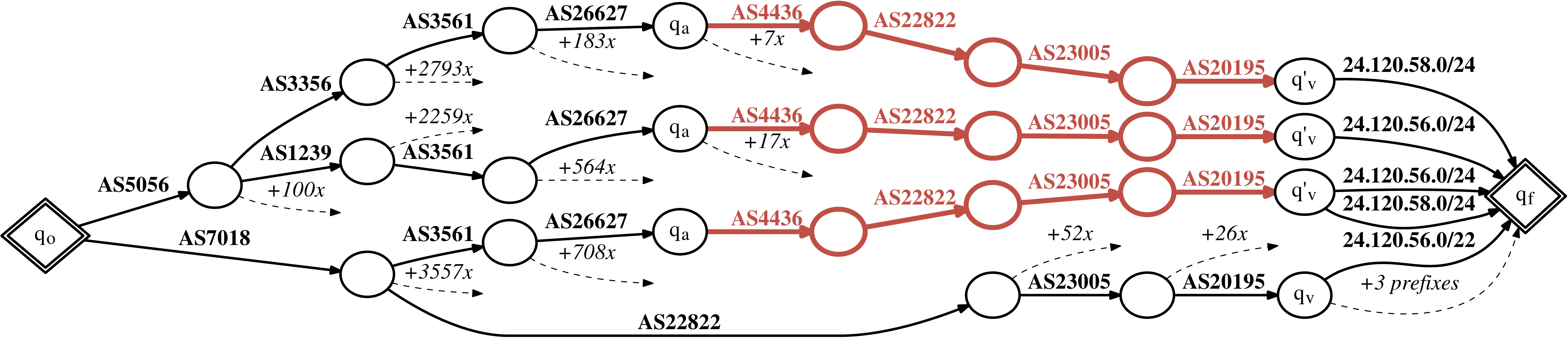}
 \caption{Route automaton for the DEFCON Attack (\texttt{AS20195}). August 10, 2008.}
 \label{fig:cair_attack_as20195}
\end{figure*}

\autoref{fig:cair_attack_as20195} shows the resulting route automaton that
corresponds to the alert raised by CAIR. The victim {\texttt{AS20195}}
(\textit{Sparkplug}) initially advertises four distinct prefixes including
{\small\texttt{24.120.56.0/22}}. During the attack, two more specific prefixes
({\small\texttt{24.120.56.0/24}} and {\small\texttt{24.120.58.0/24}}) appear to
originate from this AS. With our automaton, we can identify the highlighted
artificial path segments. Note that none of the corresponding states shows any
other outgoing transitions in contrast to the valid segment. We can locate the
attacker at $q_a$: it is either {\texttt{AS26627}}, or a common customer of this AS
and {\texttt{AS4436}} (see \autoref{sec:cair:detection:loc}). Our
observations fully comply with the public, verified information about that
DEFCON incident~\cite{blackhat}. The attacker was located at
{\texttt{AS4436}}.

%
%
%
%
%

\paragraph*{The Tehran Incident (2013)}

Another interception alert of particular interest that was reported by CAIR
affects {\texttt{AS25306}} (\textit{INSTITUTE-ISIRAN IsIran}). We refer to
this event as the \textit{Tehran Incident}. Compared to the DEFCON attack, the
analysis of this incident is more challenging for two reasons. First, this
incident has not been discussed in the public so far. Second, the attack took
place in 2013, which makes verification based on additional data sets
difficult.

\begin{figure*}[t!]
 \centering
 \includegraphics[width=\textwidth]{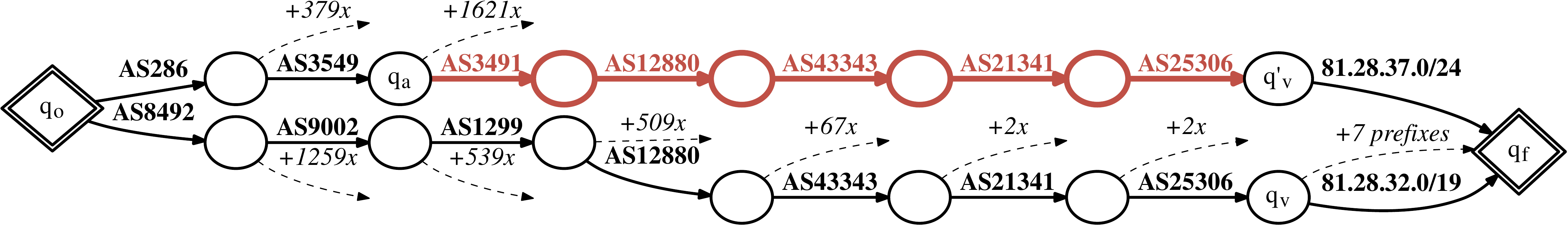}
 \caption{Route automaton for the Tehran Incident (\texttt{AS25306}). August 15, 2013.}
 \label{fig:cair_attack_as25306}
\end{figure*}


\autoref{fig:cair_attack_as25306} shows the corresponding details. On August
15, 2013, {\texttt{AS25306}} advertised eight prefixes including
{\texttt{81.28.23.0/19}}. At the same time, we observe an artificial path $w^a_v \in \Sigma_{AS}^*$ of
length 5 to the subprefix {\texttt{81.28.37.0/24}}. Within our route automaton
we are able to localize the attacker at the state $q_a$. Our technique yields
that he must be a common customer of both {\texttt{AS3549}} (\textit{Level3}) and
{\texttt{AS3491}} (\textit{Beyond the Network America (BTNA)}).

To substantiate our observation, we searched operator mailing lists and
online discussion platforms for further evidence. The incident itself was not reported. However,
we may speculate about potential customers of the ISPs involved, who maybe benefit from
an interception. Level~3 is a \textit{tier~1} ISP with a quite diverse set of
customers. BTNA seems to have a particular reputation as a
spammer-friendly service provider. Note that spam activities also
misuse BGP~\cite{rf-unlbs-06} to prevent backtracking. Although interception
itself has not been documented so far in the context of spamming, an operator
of a spamming network could be interested to implement this attack: Any
mitigation mechanism at the receiver side that requires the correct end point
(e.g., callback verification) can be fouled using the backhaul path.
Furthermore, we found an abuse report~\cite{nanog-102010}, where BTNA
apparently served as upstream for a hijacking attack during an other event.

To conclude, we do not have a proof that the Tehran Incident was indeed an
interception attack. However, based on manual investigation we found strong
evidence that the alarm triggered by CAIR was correct.

\subsection{Analysis of Route Leaks}

To further demonstrate the expressiveness of CAIR, we derive a simple yet
effective metric to assess route leaks. In its most basic form, a route leak is
given by a multi-homed AS, in the following called the \textit{originator}, who
(accidentially) re-advertises its full routing table into BGP. Upstream ISPs
often prefer customer routes~\cite{RFC-4264} and thus redistribute the
corresponding BGP announcements. The result is a major shift in global Internet
traffic such that packet flows are now redirected towards the originator.
Although a route leak leads to significant activity in BGP, it leaves
individual AS links intact. As a consequence, network graphs are ill-suited for
the detection and analysis of such events.

\paragraph*{A measure for routing dominance}

In order to asses the routing changes imposed by route leaks, we introduce a
metric for \textit{routing dominance}. Recall that each state $q \in Q$ in our
minimized automaton represents an equivalence class, i.e.~a set of partial
routes $\vec{\mathcal{L}}(q)$ accepted by $q$ (see
\autoref{sec:cair:automata}). We extend this abstract concept to ASes $u \in
\Sigma_{AS}$ such that $\vec{Q}(u)$ yields the set of all states that are
reachable via transitions labeled with $u$ as given by
$$\vec{Q}(u) = \{ q \in Q \ | \ \delta^*(\vec{q}, uw) = q, \ \vec{q} \in Q \setminus \{q_f\}, \ w \in \Sigma_{AS}^* \}.$$
The size of $|\vec{Q}(u)|$ implicitly represents the number of (partial) routes
over AS $u$. The larger this set is for a given AS, the more of its advertised
routes are redistributed by its peers, i.e.~the more it dominates routing in
BGP. Contrary to intuition, this metric does not yield a particular high rank
for the peers of our BGP collector. They exhibit an average rank of 3,808. The
top-5 ASes ranked by $|\vec{Q}|$ are \texttt{AS2914} (\textit{NTT}),
\texttt{AS3356} (\textit{Level3}), \texttt{AS3257} (\textit{Tinet}),
\texttt{AS1299} (\textit{Telia}), and \texttt{AS174} (\textit{Cogent}), which
all are prominent \textit{tier-1} providers.

\paragraph*{Regular route updates and route leaks}\label{sec:eval:leaks}

We evaluate regular changes in $|\vec{Q}|$ for individual ASes over seven years
based on our RouteViews data set and compare the results to a recent route leak
caused by \texttt{AS4788} (\textit{Telekom Malaysia}) on June 12, 2015 between
08:00 and 10:00. On average, we observe 9,019 ASes that experience any change
in $|\vec{Q}|$ during the periods of two weeks. Note that the average number of
ASes that newly appear in BGP---and thus regularly change $|\vec{Q}|$ for their
upstream ISPs---is 364, i.e.~significantly lower. Taking into account this
growth in BGP participants, we obtain a corresponding average of 19.61\% of
ASes with changes in $|\vec{Q}|$. For the Malaysia route leak, we observed a
change in routing dominance for 7,498 ASes (14.67\%) in an interval of less
than two hours.

We further observe an average of 76,785 states that are added to or removed
from the set $\vec{Q}(u)$ of any AS $u \in \Sigma_{AS}$. In contrast to that,
the Malaysia route leak led to a total change in reachability of 248,760
states. Such a significant and abrupt re-routing is unprecedented during normal
operations. 
The largest (regular) individual increase of 54,402 newly reachable states
attributes to \texttt{AS3257} (\textit{Tinet}). For \texttt{AS2914}
(\textit{NTT}), we find the largest decrease of 39,442~states. During the route
leak, in contrast, we can observe that a single AS, namely \texttt{AS577}
(\textit{Bell Canada}) loses reachability of 74,735~states (-98.57\%) within
the duration of the event. At the same time, the size of $|\vec{Q}|$ for
Telekom Malaysia increased by 32,621~entries (+2,239\%). Such a sudden
significant increase in global routing dominance is unlikely the result of a
regular change in peering agreements.


\paragraph*{The Malaysia Route Leak (2015)}

\begin{table}[t!]
  \small
  \centering
  \setlength{\tabcolsep}{3.95pt}
  \begin{tabularx}{0.475\textwidth}{r|X|r|rr|r}
    \multicolumn{2}{c}{} & \multicolumn{1}{c}{} & \multicolumn{2}{c}{\ssmall\bf ongoing route leak} & \multicolumn{1}{c}{} \\
    \multicolumn{2}{c}{Top-10 ASes} & \multicolumn{1}{c}{\textcolor{egray}{\ssmall \textbf{06:00~($T_1$)}}} & \multicolumn{2}{c}{\ssmall \textbf{08:00~($T_2$) -- 10:00~($T_3$)}} & \multicolumn{1}{c}{\textcolor{egray}{\ssmall \textbf{12:00~($T_4$)}}}\\
    \toprule
    \multicolumn{1}{c|}{} & \multicolumn{1}{c}{$u \in \Sigma_{AS}$} & \multicolumn{1}{|c}{\textcolor{egray}{$|\vec{Q}_{T_1}(u)|$}} & \multicolumn{2}{|c}{$|\vec{Q}_{T_3}(u)| - |\vec{Q}_{T_2}(u)|$} & \multicolumn{1}{|c}{\textcolor{egray}{$|\vec{Q}_{T_4}(u)|$}} \\
    \multicolumn{6}{c}{}\\[-6pt]
    \toprule
    \multicolumn{1}{r|}{\bf 1.} & AS577 & \textcolor{egray}{75,811}& -74,735 & (\textit{-98.57\%}) & \textcolor{egray}{75,660}\\
    \multicolumn{1}{r|}{\bf 2.} & \underline{AS4788} & \textcolor{egray}{\bf 1,555} & +\textbf{32,621} & \hspace{-10pt}(+\textbf{2,239\%}) & \bf \textcolor{egray}{1,412} \\
    \multicolumn{1}{r|}{\bf 3.} & AS1267 & \textcolor{egray}{32,514} & -32,218 & (\textit{-99.11\%}) & \textcolor{egray}{32,541} \\
    \multicolumn{1}{r|}{\bf 4.} & AS174 & \textcolor{egray}{94,702} & -16,770 & (\textit{-17.73\%}) & \textcolor{egray}{93,905} \\
    \multicolumn{1}{r|}{\bf 5.} & \underline{AS3549} & \textcolor{egray}{\bf 16,455} & +\textbf{11,523} & (+\textbf{71.1\%}) & \bf \textcolor{egray}{13,359} \\
    \midrule
    \multicolumn{1}{r|}{\bf 6.} & AS3356 & \textcolor{egray}{101,547} & -4,522 & (\textit{-4.45\%}) & \textcolor{egray}{101,262} \\
    \multicolumn{1}{r|}{\bf 7.} & AS6453 & \textcolor{egray}{72,213} & -3,965 & (\textit{-5.48\%}) & \textcolor{egray}{71,918} \\
    \multicolumn{1}{r|}{\bf 8.} & \underline{AS6695} & \textcolor{egray}{\bf 713} & +\textbf{3,782} & (+\textbf{531\%}) & \bf \textcolor{egray}{719}\\
    \multicolumn{1}{r|}{\bf 9.} & AS2914 & \textcolor{egray}{107,226} & -3,733 & (\textit{-3.48\%}) & \textcolor{egray}{106,862} \\
    \multicolumn{1}{r|}{\bf 10.} & AS1299 & \textcolor{egray}{117,673} & -3,416 & (\textit{-2.90\%}) & \textcolor{egray}{117,387} \\
    \bottomrule
    \multicolumn{6}{c}{}\\[-7pt]
    \multicolumn{6}{l}{\ssmall \textbf{Most affected:} \texttt{AS577} Bell, \texttt{AS1267} Wind, \texttt{AS174} Cogent} \\
    \multicolumn{6}{l}{\ssmall \textbf{Propagators: \,} \texttt{AS4788} Tel. Malaysia, \texttt{AS3549} Level3, \texttt{AS6695} DE-CIX} \\
  \end{tabularx}
  \caption{ASes by amount of newly \mbox{(un-)reachable} states. Malaysia Route Leak, June 12, 2015.}
  \vspace{-10pt}
  \label{table:cair:leak_rl}
\end{table}

The incident started on June 12, 2015, at 08:40, and began to gradually settle
down about two hours later.
We split the event in four intervals.  For $T_2$--$T_3$ (08:00--10:00), we
evaluate the top-10 ASes that exhibit the highest absolute change in reachable
states $|\vec{Q}|$. We compare our results to the as-is state before and after
the event, i.e.~at $T_1$ (06:00) and $T_4$ (12:00). Note that the overall size
of the automata hardly changed during the time frame of the route leak (-3.96\%
of states and -10.47\% of transitions).  \autoref{table:cair:leak_rl} shows
further results.


We already observed a remarkably high shift in globally dominating routes that
lies well above the average during normal operations. We see a vast increase in
reachable states $|\vec{Q}|$---which notably correlates to inbound
traffic---for the originator \texttt{AS4788} and correspondingly for his
upstream ISPs \texttt{AS3549} (\textit{Level3}) and \texttt{AS6695}
(\textit{DE-CIX}). Note that the latter does not provide upstream connectivity
under normal circumstances. Its high increase in dominant routes is rather an
artifact: Telekom Malaysia peers with the public route servers at DE-CIX, which
redistribute routes from 456 connected ASes to a total of 14,542 different
ASes. Apparently, \texttt{AS4788} leaked these routes such that the AS path of
corresponding BGP updates comprised \texttt{AS6695}. This is surprising since
route servers at DE-CIX Internet Exchange Point operate transparently and
effectively hide their own AS (see \autoref{sec:background:interception}).

The largest decrease in $|\vec{Q}|$---both in relative and absolute
terms---attributes to \texttt{AS577}. A similar observation can be made for
\texttt{AS1267} (\textit{Wind Telecomunicazioni}) and also to some extent for
\texttt{AS174} (\textit{Cogent}). Since the originator's routes are globally
prefered during the incident, these ASes consequently lose a major part of their
inbound traffic in exchange for an increase in outbound traffic towards
\texttt{AS4788}. Note that only a smaller number of all ASes (14.67\%)
propagate routes towards \texttt{AS47888}; the greater part (85.33\%) is
affected outbound, i.e.~only receives corresponding routes. Within two hours
after the event, routing converged back to its original state (see right column
in \autoref{table:cair:leak_rl}).

\paragraph*{Towards a reliable detection scheme}

With CAIR, we are able to precisely identify the role of each party involved in
a route leak. We already showed that only a moderate number of ASes is directly
affected by re-routing, which in turn can be classified into groups that either
gain or lose in reachable states $|\vec{Q}|$. The highest absolute increase in
reachable states attributes to the originator of a leak, followed by his
upstream ISPs. Hence, we are able to identify \emph{catalyst} upstreams, which
unwittingly propagate the leak. Note that the measure $|\vec{Q}|$ directly
correlates to the amount of traffic attracted by an AS. As a consequene, we can
further assess the impact of a route leak on individual ISPs.

Based on these observations, we can easily derive a threshold to detect
emerging route leaks.  We will analyze more incidents in our future work.


\section{Practical Aspects of\hspace{5pt}{\texttt{\fontsize{14}{16}\selectfont CAIR}}}\label{sec:practical}

For large volumes of input data such as a global set of routes in BGP, the
construction of route automata is challenging. In the following, we describe
our approach to minimization in CAIR in detail.
and thoroughly quantify the performance of CAIR. We show that
the required resources for the route automaton are competitive with graphs and
can even decrease with more routes being added as an effect of minimization. We
further study the correlation between automata size and Internet growth.

\subsection{Implementation}

The most basic algorithms to implement DFA minimization have a complexity of up to
$\mathcal{O}(|\Sigma_{AS} \cup \Pi| \cdot |Q|^2)$ (for notations see
\autoref{sec:cair:automata}). Although more efficient algorithms
exist~\cite{watson:mini}, the size of our intended input data greatly exceeds
that of common use cases in language processing, both in terms of the number
of words (i.e.~observable routes) and the size of the alphabet (i.e.~nodes in
the network). For
comparison, the English alphabet consists of 26~letters, whereas the technical
size of a routing alphabet is $|\Sigma_{AS}| + |\Pi| = 2^{32} + 2^{32}$ for the
IPv4 address space of the Internet.

\subsubsection{Minimization of Automata}

To construct an MDFA efficiently, we adopt a special-purpose
algorithm~\cite{daciuk:mafsa} that minimizes a DFA \textit{during}
construction without ever holding the full non-minimized automaton in memory.
Its memory complexity is $\mathcal{O}(|Q_m|)$, where $|Q_m|$ is the number of
states in the \textit{minimized} automaton.  The algorithm is capable to
randomly add or remove routes, whereas common minimization algorithms need to
re-minimize at each change of data.

Note that this particular approach expects an acyclic transition function
$\delta^*$ such that
\vspace{5pt}
$$\forall r \in \mathcal{L}(M) \ : \ \nexists \ q \in Q \ : \ \delta^*(q,r) \ = \ q \ .\\[4pt]$$
Hence, corresponding automata do not support languages where symbols or
substrings repeatedly occur within a given string. In the next section, we
show that this is not a limitation in our context.

\subsubsection{Incremental Construction Algorithm}

The procedure to add a route to a given (possibly empty) route automaton is
inspired by~\cite{daciuk:mafsa} and comprises three major steps. First, for
each route the automaton is traversed along the longest common path segment,
thereby ensuring that no invalid paths are introduced. Second, states that
accept the remaining part of the route are newly created. The final step is to
carry out an on-the-fly minimization while traversing the automaton in backward
direction. In the following, we present an in-depth description of the
individual steps. Note that during construction, we utilize a register of
states $Q_R$, which is implemented as a hash table. Its keys represent unique
right languages $\vec{\mathcal{L}}(q)$ for individual states $q \in Q$. This
provides an efficient way to search for equivalent states, i.e.~for states with
identical right languages.

\paragraph*{Step~1 \, \textit{Common Path Traversal}} To add a new route $r \in
\mathcal{L}$ to $M$, we start at $q_0$ and traverse the automaton along a
(possibly empty) sequence of existing states for the longest common path
segment $w_{lp}$ in $M$. For so-called \textit{confluence states} that have
more than one incoming transition, we need to create a new state with identical
outgoing transitions and link it to the last state traversed. This cloning
process prevents inadvertently adding false paths. Note that this is inherently
the case with network graphs.

\paragraph*{Step~2 \, \textit{Remaining Route Insertion}} After finding a
specific state $q$ that represents the longest common path segment $w_{lp}$, it
is removed from the state register $Q_R$ since its right language is about to
change. If the whole route is accepted in the first step, i.e.~if $|w_{lp}| =
|r|$, it is already contained in the automaton. Otherwise, we add new states
for the remaining part of the route and link them with corresponding
transitions. The last created state is marked as an accepting state.

\paragraph*{Step~3 \, \textit{On-the-fly Minimization}} Finally, we traverse
the automaton in backward direction along the sequence of states that represent
the newly inserted route $r$. For each state $q$, we search our register $Q_R$
for an equivalent state $\tilde{q}$ that already exists in $M$. If found, we
discard $q$ and link its preceding state to $\tilde{q}$. Otherwise, $q$ is
unique across all states of $M$ and needs to be added to the register $Q_R$.

Note that by traversing the automaton backwards in step 3, the comparison of
right languages to identify existing equivalent states is reduced to a
comparison of transitions, since recursive application would only yield the
results of already compared states. Hence, it is sufficient for keys in $Q_R$
to represent the transitions of individual states instead of their full right
language, which greatly reduces computational complexity.

Interestingly, by leaving out step 3, we obtain \textit{trie} data structures
(see \autoref{sec:need-cair}). Beside redirection of transitions, this step
only removes states from memory. As a consequence, route automata strictly
outperform tries with respect to memory requirements and expressiveness due to
the absence of redundancy.

To summarize, our construction algorithm allows to randomly add or remove routes while still ensuring minimality. 
CAIR is thus particularly well-suited for continuous monitoring
of routing changes in BGP. We can also create and archive
individual automata that represent the global routing system at specific points
in time. We made use of this feature in our evaluation.

\paragraph*{Solving Loops}

Our finite route language, and thus CAIR, account for the construction of any
set of network routes consisting of nodes and links \cite{motamedi:survey} as
long as the input is cycle-free. Even though most routing protocols such as BGP
provide built-in support for loop prevention, loops may be included in routing
data. Network operators, for example, may decide to influence route selection
by adding their own AS number multiple times to the AS path (\emph{AS path
prepending}). Still, this is not a problem for our approach: A simple way to
model subsequent occurrences of a particular AS $o$ in $M$ is to extend
$\Sigma_{AS}$ by multiple instances $o_i \in \Sigma_{AS} \text{ \ with \ } i
\in \{1,2,\dots\}$.

\begin{table}[t!]
\small
\setlength{\tabcolsep}{4.1pt}
\renewcommand{\arraystretch}{1.14}
\begin{tabularx}{0.475\textwidth}{Xrrr}
  \multicolumn{2}{r}{\ssmall August 10, 2008\ } & \multicolumn{1}{c}{\ssmall November 1, 2015} \\
  \toprule
  \bf RIB entries \textcolor{egray}{$|\mathcal{L}_{ore}|$} & \bf 10,686,819 & \bf 22,303,775 \\
  \midrule
  IP prefixes \textcolor{egray}{$|\Pi|$} & 276,706 & 600,216 \\
  AS numbers \textcolor{egray}{$|\Sigma_{AS}|$} & 29,203 & 52,396 \\
  Unique AS paths$^1$ \textcolor{egray}{$|\{w$\hspace{0.5pt}$|$\hspace{0.5pt}$wp$\hspace{0.75pt}$\in$\hspace{0.75pt}$\mathcal{L}\}|$}& 1,421,062 & 2,875,026 \\
  \midrule
  CAIR states \textcolor{egray}{$|Q_{ore}|$} & 168,184 & 302,598 \\
  CAIR transitions \textcolor{egray}{$|\delta_{ore}|$} & 4,939,314 & 10,355,671 \\
  \midrule
  Graph nodes \textcolor{egray}{$|N_{ore}|$} & 305,909 & 652,612 \\
  Graph links \textcolor{egray}{$|L_{ore}|$} & 339,515 & 725,425 \\
  \bottomrule
\end{tabularx}
\caption{Objects created after importing RIB data.}
\label{table:cair_results:rib}
\centering
{\small$^1$Unprepended AS paths only.}
\vspace{10pt}
\end{table}

\subsection{Performance Properties of\hspace{5pt}{\texttt{\fontsize{14}{16}\selectfont CAIR}}}

\begin{figure*}[t!]
  \centering
  \subfloat[Absolute growth of data structures.]{\label{fig:cair_performance:abs}\includegraphics[width=0.321\textwidth]{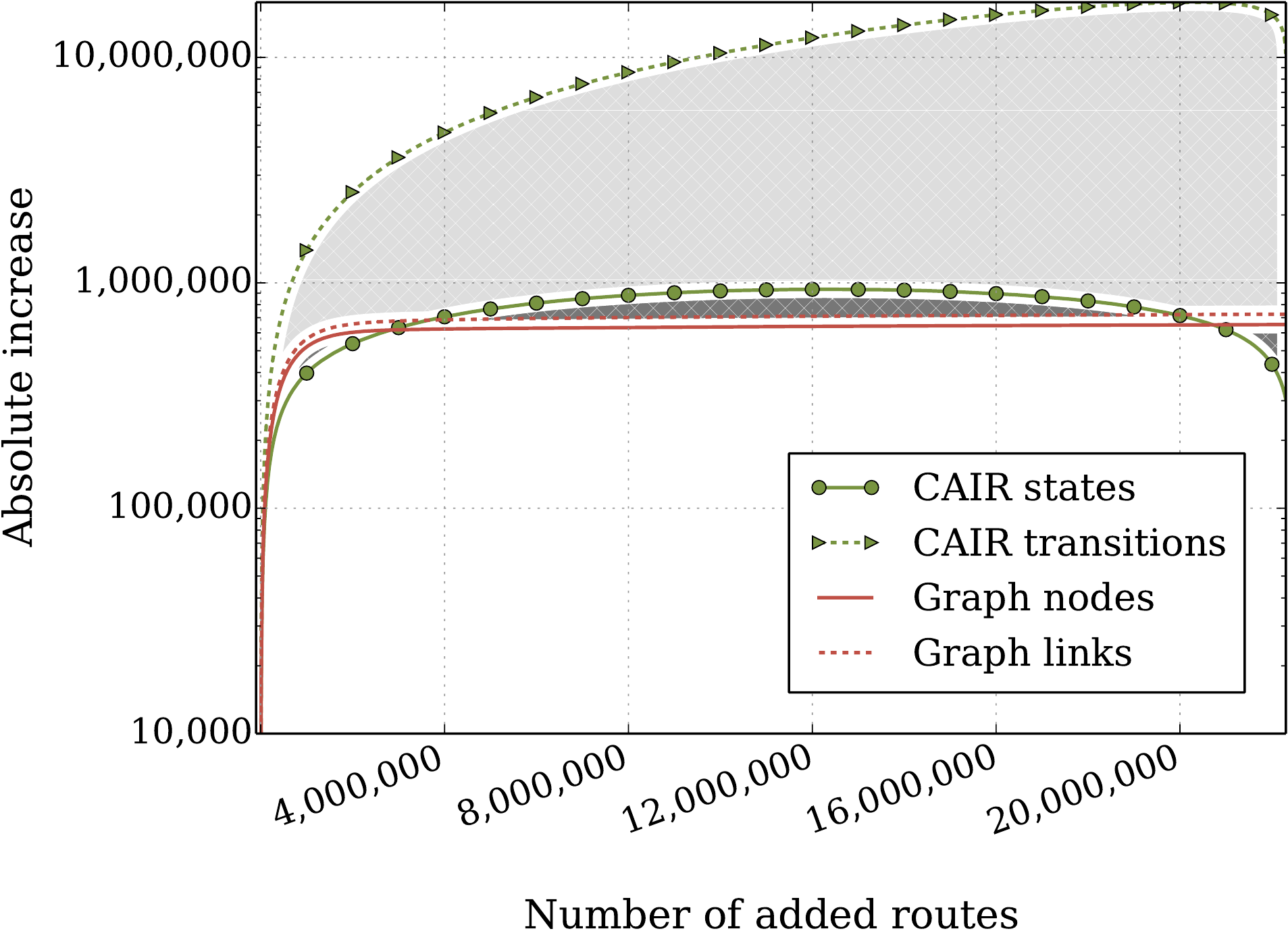}}
  \quad
  \subfloat[Relative growth of data structures.]{\label{fig:cair_performance:rel}\includegraphics[width=0.313\textwidth]{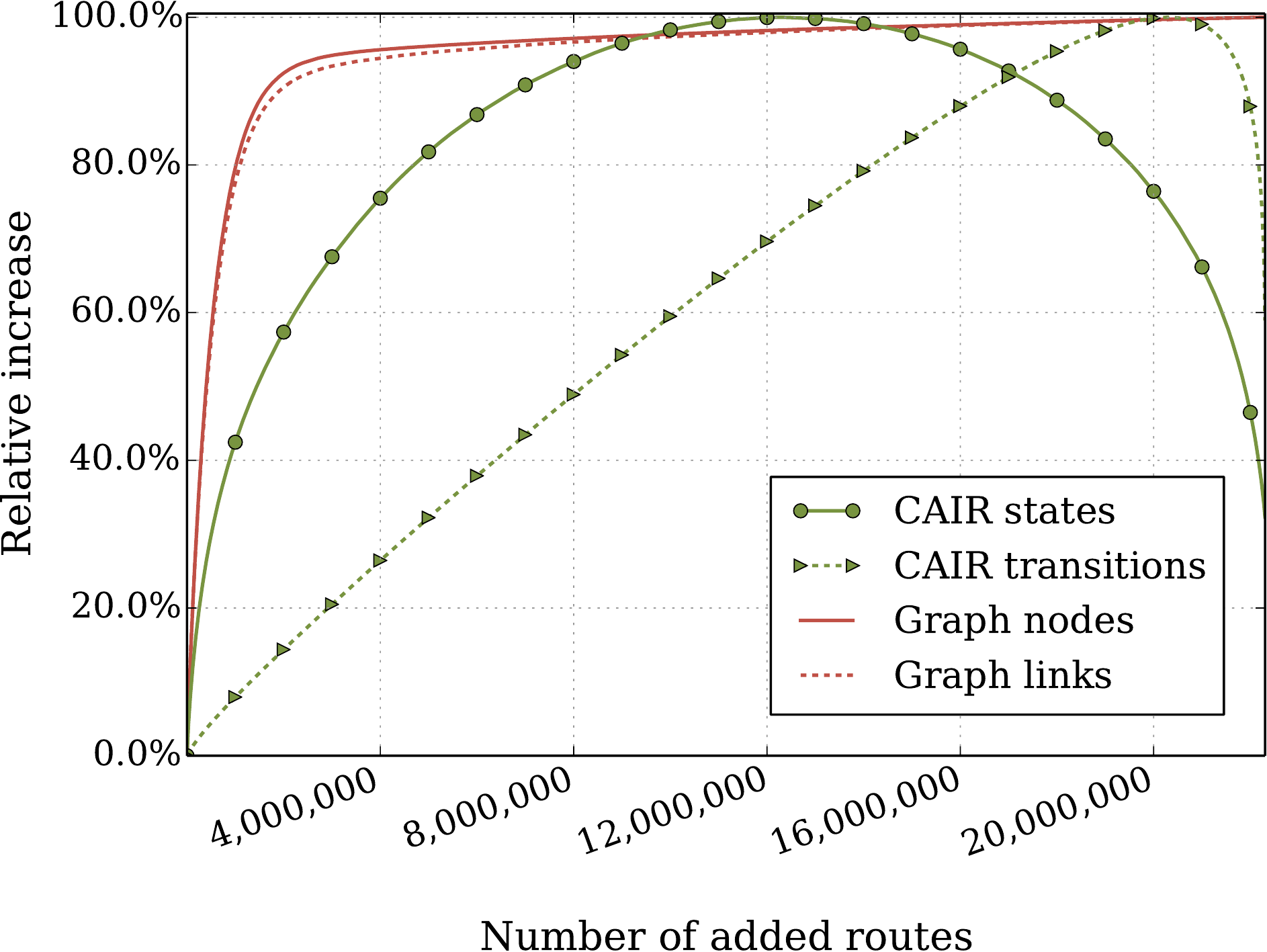}}
  \quad
  \subfloat[Information content over time.]{\label{fig:cair_evolution}\includegraphics[width=0.317\textwidth]{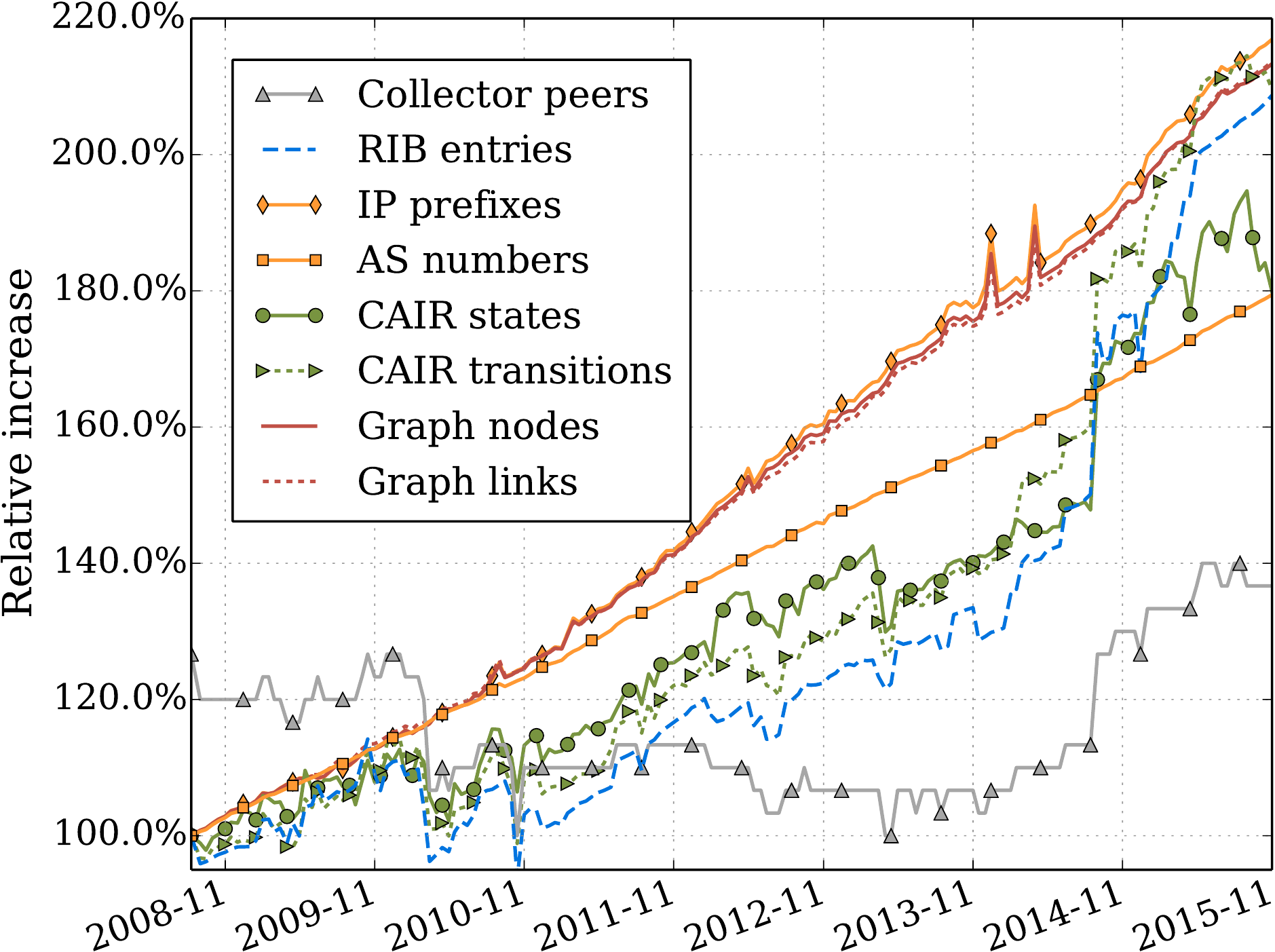}}
 \captionof{figure}{Performance characteristics of CAIR in comparison to network graphs.}
 \label{fig:cair_performance}
\end{figure*}

To study the performance of CAIR in more detail, we construct a route automaton using
the BGP routing table export from the RouteViews~\cite{routeviews} collector
\textit{Oregon2} on November 1, 2015.

\paragraph*{States and Transitions}

The RouteViews collector peers with 41~ASes that represent our set of
observation points $P_{ore} \subset \Sigma_{AS}^*$. These peers advertise
600,216 IP prefixes $p \in \Pi$ via 2,875,026 distinct AS paths $w \in
\Sigma_{AS}^*$ that comprise 52,396 individual ASes $o \in \Sigma_{AS}$. The
full set of routes known to the collector is thus given by $\mathcal{L}_{ore}
\subset \{ wp \ | \ w \in \Sigma_{AS}^*, p \in \Pi \}$ and consists of
$|\mathcal{L}_{ore}| = 22,303,775$ routes. The corresponding route automaton
$M_{ore}$ accepts the finite route language $\mathcal{L}_{ore}$. It utilizes
$|Q_{ore}| = 302,598$ states and $|\delta_{ore}| = 10,355,671$ transitions to
hold the entire RIB export (\autoref{table:cair_results:rib}). It is worth
mentioning that CAIR clearly outperforms a trie~\cite{briandais:trie} holding
the same information, which would require 73.71 times as many nodes and 2.15
times the amount of transitions.

Our vanilla implementation in Python needs 18.4 minutes to parse the input data
and 99.4 minutes to create the automaton; subsequent updates can be applied in
real-time. An optimized C++ version for operational deployment is ongoing work.

\paragraph*{Comparison to network graphs}

We define a network graph $\mathcal{G} = (N, L)$ as a set of nodes $N =
\Sigma_{AS} \cup \Pi$ and a set of links $L \subset \Sigma_{AS} \times
\Sigma_{AS} \cup \Pi$, which yields a total number of $|N| = 652,612$ nodes and
$|L| = 725,425$ links for the collector studied above. In comparison, CAIR
requires only 46.37\% of states but 14.28 times more transitions to represent
the full routing table. Taking into account that CAIR also holds all observed
routes on top of individual AS links as represented by the graph, its
efficiency is remarkably competitive.


\subfigref{fig:cair_performance:abs} quantifies the evolvement of states and
transitions in CAIR and the network graph while consecutively importing new
routes. CAIR requires fewer states but needs more transitions to implement its
expressiveness. The minimization approach in CAIR, however, introduces
self-adaptive optimization. To clarify this,
\subfigref{fig:cair_performance:rel} shows the number of states (transitions)
relatively to their maximum number during construction. While the graph data
structure created $\approx$90\% of the required objects already after importing
10\% of all routes in the input data set, CAIR grows slower. This nicely
illustrates that the graph, in contrast to CAIR, does not learn additional
information with additional routes observed. More importantly, the amounts of
states decreases in CAIR significantly after adding 55\% of the routes. This
implies that adding further routes to the automaton can actually reduce its
size, which is due to the minimization process getting more effective on larger
volumes of input data. This finding is of particular interest with respect to
possible applications of CAIR in real-time routing analysis.

\paragraph*{Resource requirements depending on Internet growth}

We now study retroactively the routing system at particular points in time. We
use the RouteViews data as described above to construct different route
automata in an interval of two weeks over the period of August, 2008 till
November, 2015. \autoref{table:cair_results:rib} presents the absolute number
of required resources. \subfigref{fig:cair_evolution} shows the relative
evolution of the RIB information content compared to resource requirements in
CAIR and the corresponding graph data structure. It is clearly visible that the
graph depends mainly on the number of IP prefixes, whereas CAIR depends on the
number of RIB entries as those provide additional routing insights.


\section{Discussion}\label{sec:discussion}

\noindent\textbf{Is CAIR limited to subprefix hijacking, which could even be protected against by RPKI? \ }
No. Even if an attacker intercepts an already announced prefix, CAIR would
detect this incident. The interception leads to a different routing policy,
i.e.~the attacked prefix is announced differently compared to all other
prefixes of the origin AS. Protection of the full AS path requires BGPsec,
which will not be deployed in the near future~\cite{g-tlsir-14}, stressing
the need for protective monitoring.

\noindent\textbf{Can CAIR detect incidents in which \emph{all} prefixes of an origin AS are simultaneously intercepted? \ }
Not now. The current search pattern is based on route diversity per AS. If
all prefixes of an AS are comprehensively intercepted as announced by the
victim, we would de facto observe a uniform redistribution. However, CAIR
stores all routing paths and allows for easy integration of additional search
patterns.

\noindent\textbf{Does CAIR depend on the selection of a specific set of observation points? \ }
No.
If an intercepted subprefix hijacking occurs, the update will be visible in the
default-free zone, similar to the victim's less specific prefixes.
However, our chances to detect the hijacking of prefixes depend on the observation of competing routes. 
For this reason, we suggest to utilize a well-connected BGP collector such as the one we used in our evaluation.

\noindent\textbf{Does CAIR require training of the data set? \ }
No. CAIR transforms routing data (either measured or artificially constructed)
into a route automaton. Reasoning is defined by search patterns, which are
based on common operational practice.

\noindent\textbf{Is the concept of FRLs self-contained? \ }
Yes. The definition of finite route languages is consistent in itself and solely
based on formal languages. We assume, though, that a
partial algebra~\cite{partial} defined over such an FRL could be embedded into
Sobrinho's routing algebra~\cite{sobrinho}. Given the existence of such a link,
we would be able to approach the theory of policy-based routing~\cite{unified}
from an experimental point of view.

\section{Related Work}

\paragraph*{Graph-based models}

Internet graphs oversimplify reality~\cite{calvert:modeling,haddadi:modelingDynamics,chengchen:recovery,rwmpb-lymmi-11,wr-itrr-13}, in particular AS-level graphs as ASes are neither atomic~\cite{mfmru-bamcr-06} nor is there a clear notion of an edge between two AS
nodes~\cite{wr-itrr-13}. CAIR is in line with these observations as our
framework does not consider an AS as a single node but in the context of its abutting paths. This
provides the flexibility to model complex policy-based routing decisions. To
improve modeling, sophisticated concepts, like multigraphs and hypergraphs~\cite{rwmpb-lymmi-11}, as well as annotated edges have been proposed. With finite route languages
and corresponding route automata, CAIR provides a natural approach to formalize and
evaluate the global routing system.

\paragraph*{Routing Policies}

The Gao Rexford model~\cite{gr-sirwg-01} is a good approximation for most AS
relationships but does not hold in general~\cite{gsg-sirp-13}. Inter-AS routing
is more complex~\cite{dimitropoulos:asrel}, e.g. by 
partial transit~\cite{glhc-icr-14}, which directly affects routing export
policies. Prefix diversity as implemented by origin ASes has been analyzed
recently by Anwar et al.~\cite{anccg-iirpw-15}. CAIR naturally
reflects these prefix-specific policies. More importantly, such observations
are consistent with our assumption that an origin AS may advertise its
prefixes differently, and that an \emph{arbitrary} AS will usually not
distinguish between prefixes of the same origin AS.
We addressed exceptions in \autoref{sec:eval:application}.
A theoretical approach for policy-based routing has been introduced with Sobrinho's
routing algebra~\cite{sobrinho}. Subsequent work
focused mainly on abstract models of routing policies~\cite{pathvector,dynamicrouting}
and their unification~\cite{metarouting,unified}.

\paragraph*{Generic Hijacking Detection}

A common way to detect and assess hijacking is based on the analysis of
\one control-plane
information~\cite{phas,misconfig,rootcause,locinstable,haystack,predict,spurious},
\two data-plane
measurements~\cite{blackholes,binary,predict2,lifeguard,trinocular,collaborative},
or \three the combination of
both~\cite{fingerprint,argus,impact,hubble,optometry,poiroot}.  CAIR
belongs to the first class. We want to emphasize that CAIR is not
solely a system for anomaly and hijacking detection, but a rigorous
framework to model and assess various aspects of routing.
Bogon route leaks have been studied in the past~\cite{bogon}.
Hira et al.~\cite{hcg-clrac-13} argued that the China Telecom incident was most likely a route leak because of missing subprefix hijacking.
We complement these findings with a study of the Telekom Malaysia incident.

\paragraph*{Interception Attacks}

The impact of interception attacks was studied in detail using a next-hop signature and simulations~\cite{study}.
No decisive evidence for the disccoverd events could be found.
In contrast to this, we verified CAIR using ground truth data, and confirmed new incidents using complementary data sources.

Current approaches~\cite{bogus} that use control-plane data to detect interceptions are challenged by the lack of a complete AS-level topology~\cite{capturing,incomplete}.
Careful tuning of heuristic parameters is necessary to yield stable detection results.
CAIR, instead, is well-suited to operate on incomplete data as it explores routing properties therein.

Real-time detection was analyzed by leveraging a light-weight, distributed active measurement scheme~\cite{hopcount}.
CAIR can be deployed for real-time analysis as well, but
does not depend on active probing and thus is a noninvasive technique.
A practical demonstration of the applicability of interception attacks was presented in~\cite{defcon}.
Complementary results from an operational point of view are provided
in~\cite{blackhat}. We evaluate CAIR using this data in our ground truth
case study.

\section{Conclusion}\label{sec:conclusion}

In this paper, we introduce CAIR, a novel formalization of Internet routing to
address a fundamental problem with network graphs. Our model is designed to
preserve route diversity, where routes from $AS_1 \rightarrow AS_2$ and $AS_2
\rightarrow AS_3$ in general do not imply a route $AS_1 \rightarrow AS_3$. It
is based on formal language theory. With so-called \textit{finite route
languages}, we provide a comprehensive formal framework in which routing
aspects can be rigorously described. To put this theoretic concept into
practical use, we propose \textit{CAIR} as an implementable equivalent that is
based on the construction of minimal deterministic automata.  Corresponding
\textit{route automata} offer unique benefits in terms of efficiency and
expressivenes. In particular, CAIR preserves route diversity and solves the
transitivity problem of graphs. We will make our fully functional reference implementation available.
CAIR can readily be used to study policy-based routing.

We applied our routing model to formalize a sophisticated man-in-the-middle
attack in BGP, which can be translated into a practical search pattern that is
implementable in our route automata. With an analysis of BGP routing tables of
more than seven years, we demonstrated great potential in using CAIR for
routing analysis. We gained insight into normal and abnormal routing changes,
studied a known route leak incident in thorough detail, and identified 22
new cases of interception attacks. We showed that the applicability of
CAIR reaches well beyond the detection of routing anomalies.

\paragraph*{Future Work}

We intend to advance our approach in several directions. The detection of
interception attacks, and hijacking in general, can still be improved, and more
incidents ought to be studied in detail. We further see a need to deploy CAIR
on a continuous basis and to monitor the global routing table for route leaks
and attacks in real-time. An optimized version of our implementation to handle live BGP streams
is in development. We plan to investigate
other research areas that utilize AS-level graphs and thus can naturally
benefit from CAIR. In principle, our framework can be applied to any policy-based routing scenario. An evaluation of corresponding analysis techniques is part
of our future work. We further intend to study the link between our formal
model and algebraic approaches to routing analysis.

\clearpage


\balance
{\footnotesize \bibliographystyle{acm}
\bibliography{bib/modeling,bib/hijacking,bib/bgp}}

\end{document}